\documentclass [osajnl, twocolumn, showpacs, superscriptaddress, 10t]{revtex4-1}

\usepackage	{graphicx}
\usepackage	{amsfonts}
\usepackage	{amssymb}
\usepackage	{array}
\usepackage	{amsmath}
\usepackage	{color}
\usepackage	{verbatim} 
\usepackage	{psfrag}
\usepackage	{mathtools}

\newcommand	{\ket}	[1]	{\left\vert  	#1 \right\rangle}
\newcommand	{\bra}	[1]	{\left\langle 	#1 \right\vert}
\newcommand	{\pjct}	[2]	{\left\vert	#1 \right\rangle\!\left\langle #2 \right\vert}

\newcommand	{\ud}		{\mathrm{d}}

\DeclareMathOperator{\Tr}{Tr}

\begin{document}

\title{Detecting the degree of macroscopic quantumness using an overlap measurement}

\author{Hyunseok~Jeong}
\affiliation{Center for Macroscopic Quantum Control, Department of Physics and Astronomy, Seoul National University, Seoul, 151-742, Korea}

\author{Changsuk~Noh}
\affiliation{Centre for Quantum Technologies, National University of Singapore, 3 Science Drive 2, Singapore 117543}

\author{Seunglee~Bae}
\affiliation{Center for Macroscopic Quantum Control, Department of Physics and Astronomy, Seoul National University, Seoul, 151-742, Korea}

\author{Dimitris~G.~Angelakis}
\affiliation{Centre for Quantum Technologies, National University of Singapore, 3 Science Drive 2, Singapore 117543}
\affiliation{School of Electronic and Computer Engineering, Technical University of Crete, Chania, Crete, Greece, 73100}

\author{Timothy~C.~Ralph}
\affiliation{Centre for Quantum Computation and Communication Technology, School of Mathematics and Physics, University of Queensland, St. Lucia, Queensland 4072, Australia}

\begin{abstract}
We investigate how to experimentally detect a recently proposed measure to quantify macroscopic quantum superpositions [Phys. Rev. Lett. {\bf 106}, 220401 (2011)], namely, ``macroscopic quantumness'' $\mathcal{I}$.
Schemes based on overlap measurements for harmonic oscillator states and for qubit states are extensively investigated. Effects of detection inefficiency and coarse-graining  are analyzed in order to assess feasibility of the schemes.
\end{abstract}


\maketitle

\section{Introduction}
Macroscopic quantum states have provoked the imaginations of many physicists since the early days of quantum mechanics \cite{Schrodinger}.
Creation and detection of macroscopic quantum superpositions are difficult but interesting tasks.
Efforts towards such implementations have been made, for example, using atomic/molecular systems \cite{MonroeCat,C60}, superconducting circuits \cite{SQUID1,SQUID2}, and optical setups \cite{Ourjoumtsev,Gao,Afek}.
One very important issue in efforts to create and detect a macroscopic superposition is to have a measurable quantity that quantifies the degree of macroscopic quantumness of a given state.
Since the initial attempt by Leggett \cite{Leggett}, there have been various
proposals and discussions  of such measures \cite{Dur,Bjork,Shimizu2002,Shimizu2005,Korsbakken,Korsbakken2,Cavalcanti,Mar,LJ2011,Flowis2012,Nim2013,Sek2014,Sek2014b,Gir2014, HJRev}.
In particular, a measure that quantifies the degree of macroscopic superposition for arbitrary harmonic oscillator systems, including mixed states, was proposed \cite{LJ2011}.
This measure, which we will call ``macroscopic quantumness'' $\mathcal{I}$, can be straightforwardly calculated if the density matrix of a state under consideration is known.
However, this is generally not the case, and given the difficulties in performing a quantum state tomography it would be useful to have an experimental method to measure directly the value of $\mathcal{I}$.
In this paper, we investigate several schemes to detect $\cal I$ using overlap measurements and discuss their experimental feasibility.

 This paper is organized as follows.
In Sec.~\ref{sect2}, we briefly review measure $\mathcal{I}$ as a general quantifier of macroscopic quantum superpositions.
We then discuss in Sec.~\ref{sect3} a general method based on an overlap measurement to observe $\mathcal{I}$ experimentally, and explain two broad categories of overlap measurements.
The next two sections, \ref{sect4} and \ref{sect5}, describe each method in detail for harmonic oscillator (continuous variable) systems and two-level (qubit) systems.
In Sec.~\ref{sect67}, we investigate the effects of experimental imperfections including coarse-graining and detection inefficiency.
A summary with some remarks is given in Sec.~\ref{sect8}.

\section{Macroscopic quantumness $\mathcal{I}$}
	\label{sect2}
The original motivation behind the proposal of $\mathcal{I}$ came from the high-frequency oscillations in the Wigner functions for the Schr\"{o}dinger-cat-like states.
The Wigner function that visualizes a quantum state in the phase space can be calculated from the characteristic function, which for a density operator $\rho$ is defined as
	\begin{equation}
		\chi(\xi) =
		\Tr \left[\, \rho\, e^{(\xi\, \hat{a}^{\dagger} - \xi^{*} \hat{a})}\, \right] ,
	\end{equation}
where  $\hat{a}$ and $\hat{a}^{\dagger}$ are the bosonic annihilation and creation operators, respectively.
The Wigner function $W(x,y)$ is defined as the Fourier transform of the characteristic function \cite{QObook},
	\begin{equation}
		W(x,y) =
		\frac{1}{\pi^2} \int\!\!\!\int \ud\,\xi_{r}\, \ud\,\xi_{i}\ \chi (\xi_{r}, \xi_{i}) 
		e^{-2\, i\, (x\, \xi_{i} - y\, \xi_{r})},
	\end{equation}
where the subscript $r$ ($i$) denotes the real (imaginary) part of $\xi$.
The frequency of a Wigner-function component along the real (imaginary) axis is $\xi_{r}$ ($\xi_{i}$) and its complex amplitude for a specific frequency $\xi$ corresponds to $\chi(\xi)$.

\begin{figure}[t]
\includegraphics[scale=0.38]{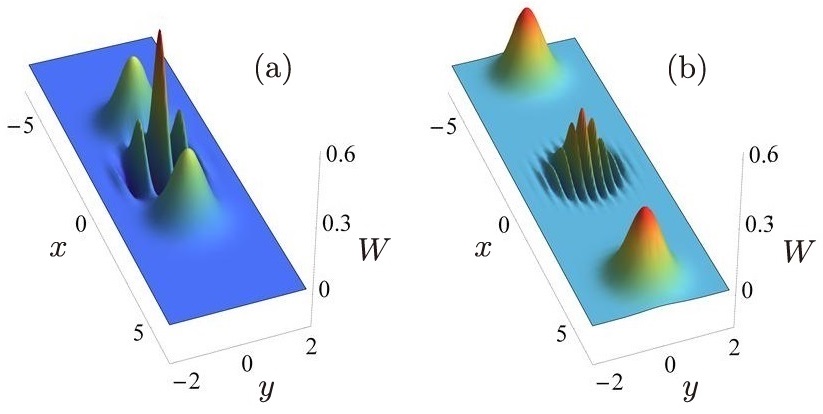}
	\caption{(Color online) Wigner functions of (a) pure superposition of coherent states $\propto \ket{\alpha} + \ket{-\alpha}$ with $\alpha=2.3$,
			and (b) partially mixed superposition of coherent states $\propto \pjct{\alpha}{\alpha} + \pjct{-\alpha}{-\alpha} + \Gamma(\pjct{\alpha}{-\alpha} + \pjct{-\alpha}{\alpha})$ with $\alpha=4.96$
			and $\Gamma\approx0.464$.
			These two cases give the same value of ``macroscopic quantumness'' ($\mathcal{I}_{\rm ho}\approx5.29$).}
\label{fig:wigner}
\end{figure}

As a typical example, let us consider a superposition of two coherent states (SCS) \cite{cat1,Ourjoumtsev,cat4,cat5,cat6}
	\begin{equation}
		\ket{\psi_{\rm{scs}}} = 
		N_{+} ( \ket{\alpha} + \ket{-\alpha} ) ,
	\end{equation}
where the amplitude $\alpha$ is assumed to be real without loss of generality and $N_{+}$ is the normalization factor.
In the phase space, its Wigner function displays interference fringes between two peaks at $\pm\alpha$ as shown in Fig.~\ref{fig:wigner}(a) for $\alpha = 2.3$.
Suppose that someone could generate a ``larger'' superposition state by increasing the amplitude $\alpha$, while the generation process makes it partly lose interference between the two coherent states.
Such a mixed state should be represented in a more general form as
	\begin{equation}
		\rho_{\rm{scs}} = 
		N_\Gamma \Big[ \pjct{\alpha}{\alpha} + \pjct{-\alpha}{-\alpha} +
		\Gamma\, (\pjct{\alpha}{-\alpha} + \pjct{-\alpha}{\alpha}) \Big],
	\end{equation}
where $0\leq|\Gamma|\leq1$ and $N_\Gamma$ is the normalization factor.
If $\Gamma=1$, it becomes a pure SCSs whereas it is totally mixed when $\Gamma=0$.
Figure~\ref{fig:wigner}(b) presents an example of a partially mixed state with $\Gamma=0.46357$ and larger amplitude $\alpha=4.96$.
Now the problem is to find a measure of macroscopic quantumness that can account for the increase in ``size" and decrease in ``quantumness" during this process.

First, we note that the ``frequency'' of the fringes (how dense the fringes are) reflects the ``effective size'' of the superposition, i.e., how far the component states separate in phase space.
Second, the ``quantum coherence'' (here we mean the degree of genuine superposition against its completely mixed version in terms of the ``pointer basis'' \cite{deco}) relates to the magnitude of the interference fringes.
The point is thus how to quantify both the ``frequency'' and ``magnitude'' of interference fringes in the Wigner representation to quantify the ``macroscopicness'' and ``quantumness'' at the same time.
One promising way is to take an integral
	$\int\! \ud^{2}\xi\, (\xi_{r}^{2} + \xi_{i}^{2})\, |\chi(\xi)|^{2}.$
This integral indeed combines the required factors, ``effective size'' and the ``degree of quantum coherence'', in a single measure.
The formal definition of the measure $\mathcal{I}_{\rm ho}$ for an $M$-mode harmonic oscillator system
\cite{LJ2011} is only slightly different:
\begin{equation}
	\label{eq:integ}
\mathcal{I}_{\rm ho} \left( \rho \right)
= \frac{1}{2 \pi^M} \!\! \int \!\! d^2 \boldsymbol{\xi}
 \sum_{m = 1}^M \! \left[ \left| \xi_m \right|^2  - 1 \right] \left| 
 \chi \left( \xi_1,\, \xi_2,\, \cdots,\, \xi_M \right) \right|^2  
\end{equation}
where $\int d^2\boldsymbol{\xi} = \int d^2\xi_1 \int d^2\xi_2 \cdots \int d^2\xi_M$ and
 $-1$ has been inserted simply to make any coherent states or their product states (regardless of their amplitudes) a reference with  $\mathcal{I}_{\rm ho}=0$.

As discussed in Ref.~\cite{Reply}, the factor $-1$ in Eq.~(\ref{eq:integ}) 
may be unnecessary as it causes $\cal I$ to have negative values for some mixed states \cite{Comment,Reply}.
If we remove the factor  $-1$ from the definition (for example, take
$\int\! \ud^{2}\xi\, (\xi_{r}^{2} + \xi_{i}^{2})\, |\chi(\xi)|^{2}$
as the definition for single-mode harmonic oscillator states), 
the measure will always be non-negative and become zero only for an extreme mixture 
$\sum^{\infty}_{n=0}|n\rangle\langle n|$ where $|n\rangle$ is a number state
 \cite{Reply}.
Since this change does not make an essential difference from the original version (\ref{eq:integ}), we stick to the original definition in this paper. 
We also note that the measure $\cal I$ is invariant under the displacement operation \cite{LJ2011} because it simply shifts the Wigner function 
without changing its shape in the phase space.
Therefore, for example, the class of states generated by applying the displacement operations
on microscopic entanglement \cite{Sek2012,Lvov2012,Bruno2012} cannot have high values of $\cal I$.

Returning to the example of SCSs, Fig.~\ref{fig:wigner}(a) corresponds to a superposition of two coherent states with a relatively small effective size but with the full quantum coherence, while Fig.~\ref{fig:wigner}(b) to the larger effective size with a partial coherence.
Sensibly, the measure $\mathcal{I}_{\rm ho}$ gives the same value ($\approx 5.29$) for the two cases. 
The aforementioned definition of $\mathcal{I}$ can be applied to arbitrary states in harmonic oscillator systems such as bosonic fields (quadrature variables of light) and mechanical states with motional degrees of freedoms.

There is also an alternative definition \cite{LJ2011} that takes note of fast decoherence rates of macroscopic superpositions \cite{deco}. Taking the purity decay rate as a measure of a macroscopic superposition,
a more general definition of $\cal I$ is
	\begin{equation}
		\mathcal{I}(\rho) =
		-\frac{1}{2} \frac{\ud\mathcal{P}(\rho)}{\ud\tau} ,
	\label{eq5}
	\end{equation}
where $\mathcal{P}(\rho) = \Tr(\rho^{2})$ is the purity of state $\rho$ and $\tau$ = (decay rate)$\times$(time) is the dimensionless time.
It is straightforward to show that $\mathcal{I}$ may also be represented as
	\begin{align}
		\mathcal{I}(\rho) =
		-\Tr \left[ \rho \mathcal{L}(\rho) \right] ,
	\end{align}
where $\mathcal{L}(\rho) = \ud\rho / \ud\tau$ is a Lindblad superoperator.
Interestingly, this seemingly unrelated definition of $\mathcal{I}$ is shown to be identical to the definition given in Eq.~(\ref{eq:integ}) \cite{LJ2011}, if one takes a well-known decoherence model:
	\begin{equation}
		\mathcal{L}_{\rm ho}(\rho) =
		\sum_{m=1}^{M} \left[\, \hat{a}_{m}\, \rho\, \hat{a}^{\dagger}_{m} -
		\frac{1}{2}\, \rho\, \hat{a}^{\dagger}_{m} \hat{a}_{m} -
		\frac{1}{2}\, \hat{a}^{\dagger}_{m} \hat{a}_{m}\, \rho\,\right] ,
	\label{dmodel}
	\end{equation}
which describes the decay (photon loss) mechanism for optical fields \cite{Louisell, QObook}.

Following the definition (\ref{eq5}), instead of  harmonic oscillator systems, one may also consider the degree of macroscopic quantumness for qubit states ($N$ two-level systems spanned in a $2^{\otimes N}$ Hilbert space) such as spin-1/2 systems.
Assuming that dephasing dominates over decay of multi-qubit systems, $\mathcal{L}(\rho)$ may be replaced with the dephasing model \cite{WM2009}
	\begin{equation}
		\mathcal{L}_{\rm{qb}}(\rho) =
		\frac{1}{2} \sum_{m=1}^{M}  \hat{\sigma}_{z,m}\, \rho\, \hat{\sigma}_{z,m} - \rho 
	\label{dmodel2}
	\end{equation}
and $\mathcal{I}_{\rm qb}(\rho) = -\Tr \left[ \rho \mathcal{L}_{\rm qb}(\rho) \right]$.
In this paper, we consider {\it how to measure the purity decay rate for both optical fields based on Eq.~(\ref{dmodel}) and qubit systems based on Eq.~(\ref{dmodel2})}. In what follows, we shall use notation ${\cal I}(\rho) $ without the subscripts used above as far as its meaning is obvious in the context.

\section{Detecting macroscopic quantumness $\mathcal{I}$}
	\label{sect3}

\subsection{General scheme using an overlap measurement and added decoherence}
\begin{figure}[t]
\includegraphics[scale=0.45]{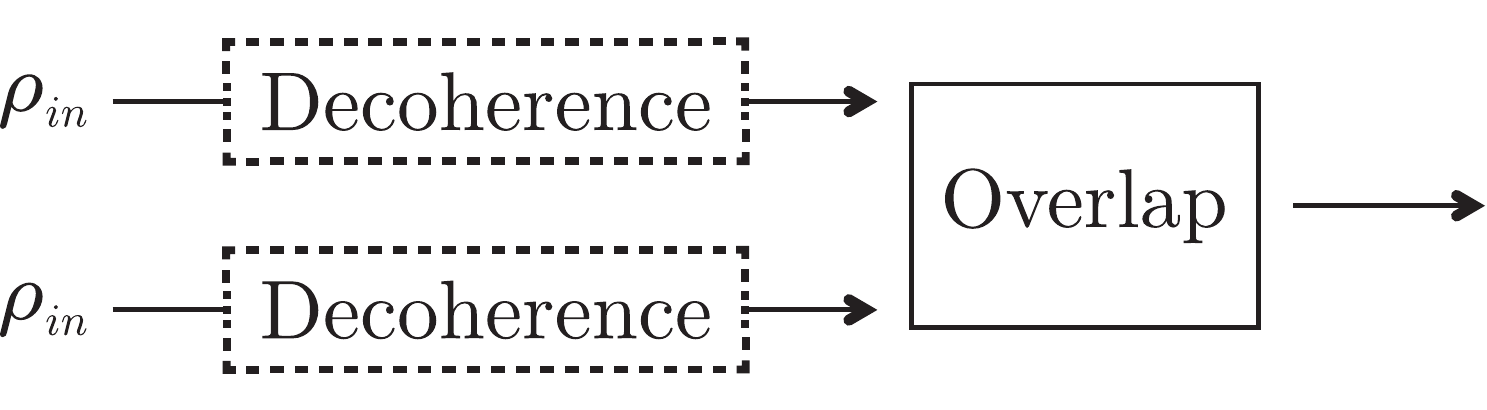}
	\caption{Measurement scheme for determining the macroscopic quantumness $\mathcal{I}$ for a density matrix $\rho_{in}$.
			The overlap is measured twice, with and without induced decoherence.}
\label{scheme}
\end{figure}

Being a nonlinear functional of $\rho$, experimental detection of $\mathcal{I}$ seems to require a reconstruction of $\rho$, which becomes quickly intractable with increasing system size.
There are, however, ways to measure nonlinear functionals of a density matrix experimentally.
In particular, given two copies of $\rho$, it is possible to measure the purity of a quantum state
\cite{swap1,swap2,swap3,swap4,swap5,swap6,swap7,added_01,added_02,SML2013}.
This fact allows one to measure $\mathcal{I}$ by simply noting that it can be written as the limit $\Delta\tau \rightarrow 0$ of 
	\begin{equation}
		\mathcal{I}_{\Delta\tau}(\rho) =
		-\frac{\mathcal{P}(\,\rho\,(\Delta\tau))- \mathcal{P}(\rho)}{2\,\Delta\tau},
	\label{coarse-grained MQ}
	\end{equation}
which is a measurable quantity, given a purity measurement scheme and ways to induce small decoherence. The situation is depicted in Fig.~\ref{scheme}.

For optical fields, the decoherence caused by loss of photons can be induced simply by using a beam splitter of an appropriate reflectivity because of the equivalence between the two processes \cite{Leonhardt,kim1995}.
The dotted boxes in Fig.~\ref{scheme} can thus be provided by beam splitters with an appropriate ratio determined by the decoherence time $\Delta\tau$.
For the dephasing of qubit states, the method of decoherence may depend on how the qubits are implemented.
For example, in the case of polarization systems, random phases can be artificially added using wave plates in a stochastic way to implement the decoherence effects.

\subsection{Direct measurement and controlled-swap schemes for overlap measurements}

\begin{figure}[t]
\includegraphics[scale=0.48]{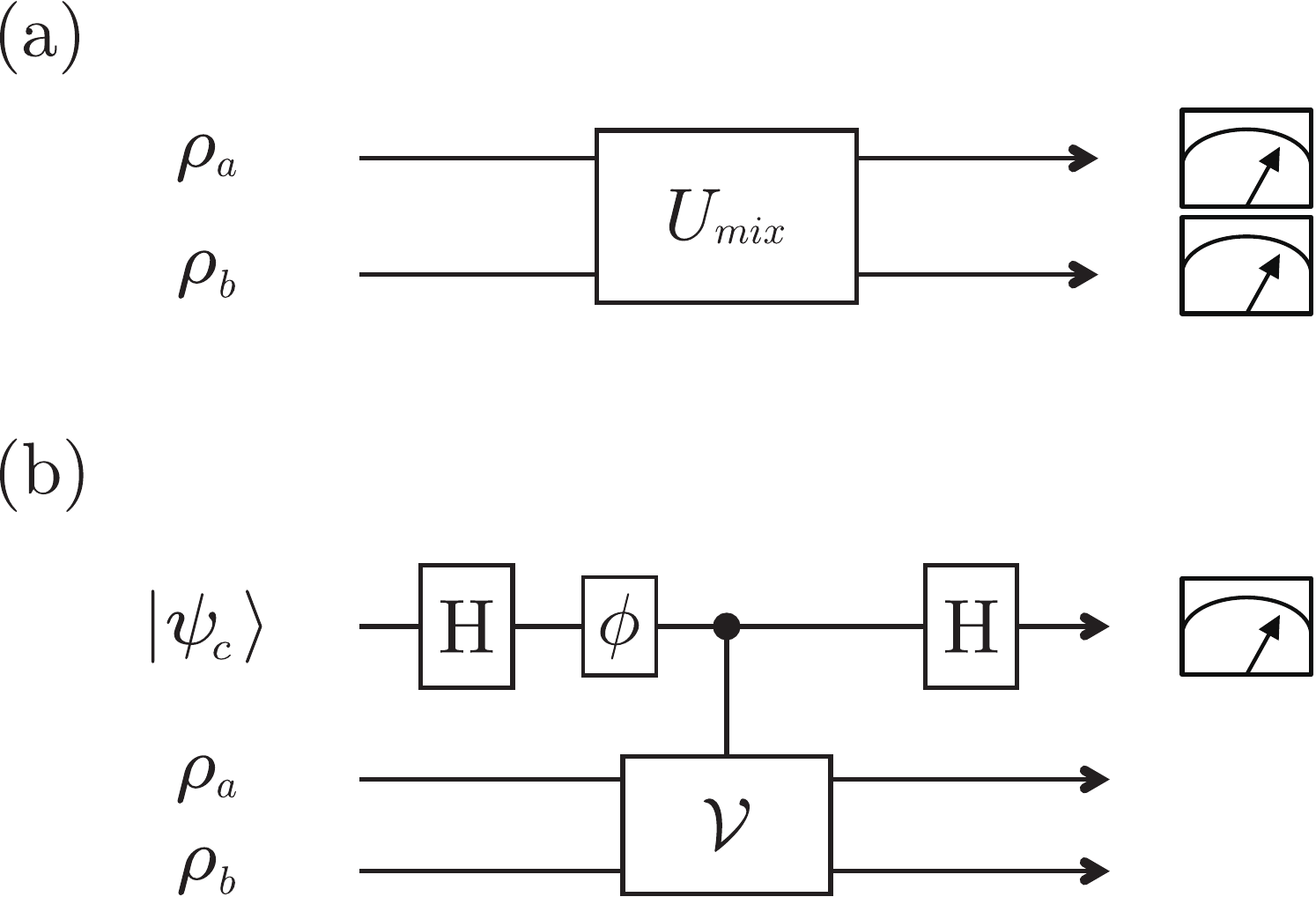}
	\caption{Two schemes of an overlap measurement.
			(a) Direct measurement of the swap operator using an unitary mixing and detections.
			(b) Controlled-swap method  where only a single detection is needed. In addition to the detector and the swap gate ($\cal V$), two Hadamard gates (H) and a phase shifter ($\phi$) are used.
			}
\label{overlap}
\end{figure}

Taking for granted that decoherence can be induced controllably via aforementioned methods, we discuss general schemes to measure the purity.
Known approaches are based on a simple relationship for an overlap between two states, $\rho_a$ and $\rho_b$:
	$\Tr [\, {\cal V} \rho_{a}\otimes\rho_{b}] = \Tr [\rho_{a}\rho_{b}]$,
where $\cal V$ is the swap operator defined by
	${\cal V}\ket{\psi_{1}}\ket{\psi_{2}} = \ket{\psi_{2}}\ket{\psi_{1}}$
\cite{swap1,swap2,swap3,swap4,swap5,swap6,swap7,SML2013}.
When $\rho_a = \rho_b = \rho$, the overlap yields the purity, so we discuss overlap measurements in order to keep the discussion more general.
Broadly speaking, the overlap measurement approaches can be divided into two categories depicted in Fig.~\ref{overlap}.

The first method, shown in Fig.~\ref{overlap}(a) is to perform a direct measurement of the unitary operator $\mathcal{V}$ \cite{swap4,swap6,swap7,Daley}, comprised of an unitary operation that mixes two input states and subsequent detections.
For the overlap between two qubits, this is equivalent to a measurement of the antisymmetric projector, which can be performed with a beam splitter and photodetectors for polarization qubits \cite{swap4}.
The case of two multi-qubit states has been studied by Alves and Jaksch \cite{swap6} in an optical lattice setting, where two internal states of the trapped cold atoms represent the qubit.
The case of harmonic oscillator states has been first addressed by Pregnell \cite{swap7}, where the author considers `decomposing' the swap operator into a unitary transformation and a measurement.
Very recently, Daley \textit{et al.}, rediscovered the method independently and generalized it to arbitrary multi-partite states \cite{Daley}.

The second method shown in Fig.~\ref{overlap}(b) is to perform a controlled-swap (C-SWAP) operation, where the states are swapped only if the control qubit is in one of the computational basis states (for example, the vertical polarization state $\ket{V}$ for a single photon polarization qubit). The scheme is similar to the usual interferometer: the qubit goes through the Hadamard gate, a phase shifter, another Hadamard gate, followed by a (visibility) measurement, but different in that after the phase shift, controlled-SWAP operation is performed. The visibility then corresponds to the overlap between $\rho_a$ and $\rho_b$ \cite{swap1,swap2,swap5}.
Advantages of this method are 1) the detection is carried out only on the control qubit and 2) it is easily generalized to an arbitrary unitary operation other than the considered swap operation.
A controlled-unitary operation can be performed with the help of pre-arranged entangled states and linear optics elements \cite{ZQ2011}.

\section{Direct measurement schemes}
	\label{sect4}

\begin{figure}[t]
	\includegraphics[width=0.3\textwidth]{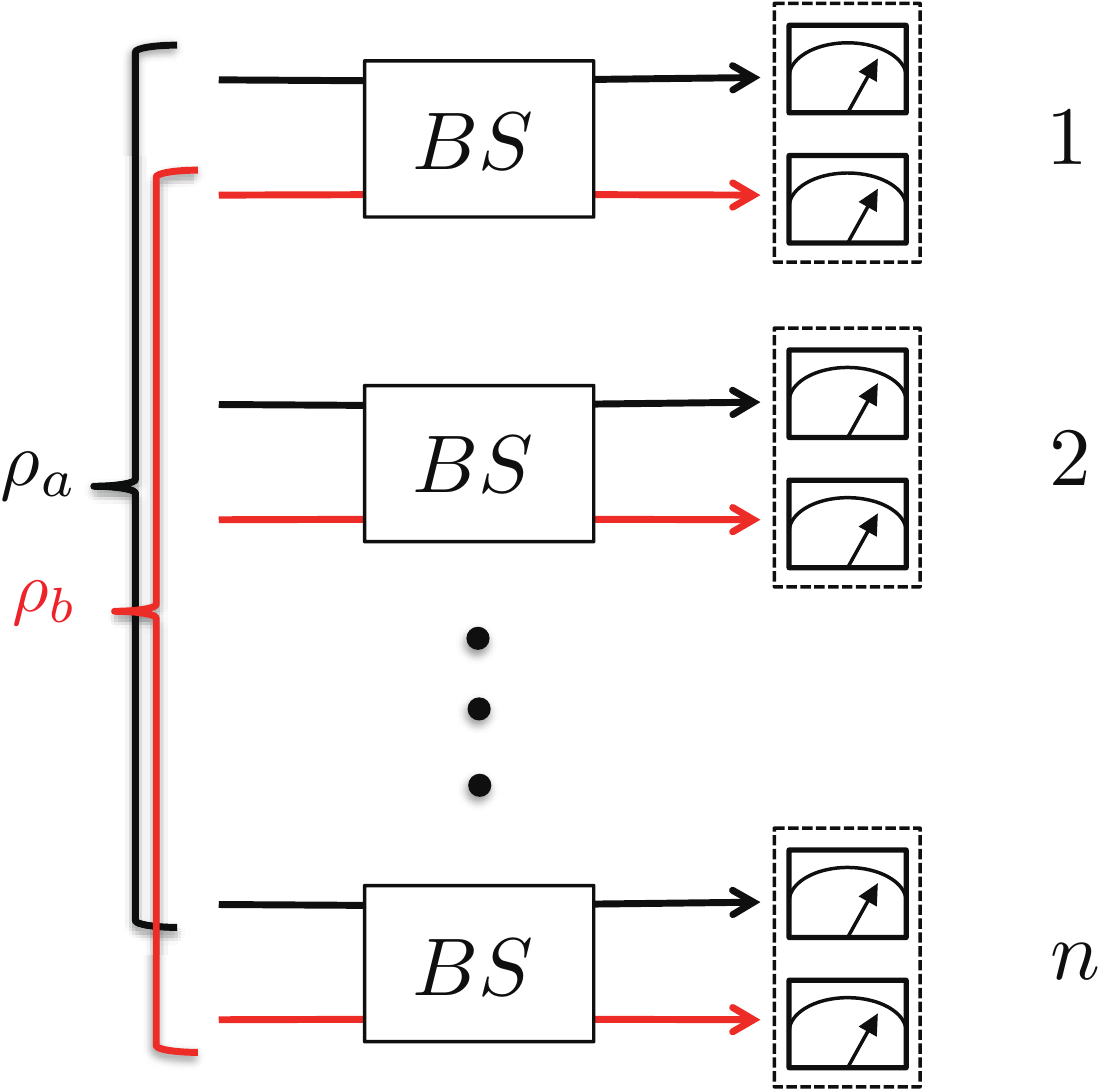}
	\caption{(Color online) Overlap measurement of $n$-partite states.
			On each site denoted $1, 2, \cdots, n$, one unit from each copy is mixed through a unitary operation and appropriate measurements are performed.
			The unitary operation denoted by $BS$ is usually the beam splitter unitary or a variant of it.
			Dotted boxes denote basic units of detection.}
	\label{npartite}
\end{figure}

We first discuss direct measurement schemes in detail for $n-$partite states.
When the system is comprised of $n$ fundamental units (in this work we take a unit to be either a qubit or a harmonic oscillator), the swap operator for two $n-$partite states is simply
	\begin{equation}
		{\cal V} = {\cal V}_{1}\otimes {\cal V}_{2}\otimes {\cal V}_{3}\otimes \cdots \otimes {\cal V}_{n} \equiv \otimes^n_{k=1} {\cal V}_{k} ,
	\end{equation}
	where ${\cal V}_k$ is a partial swap operator for the $k$-th system.
A schematic illustration of direct measurement schemes for $n$-partite states is given in Fig.~\ref{npartite}.
One unit from the first system and a corresponding unit from the second goes through appropriate unitary interaction and detection processes.
The unitary operation is of the beam-splitter type and together with detection is repeated throughout all units. In this section, we discuss these two processes in detail for harmonic oscillator states and qubit states.

\subsection{Harmonic oscillator states}
	\label{sect4a}
	
For the harmonic oscillator states, we give a brief review on the method of direct measurements discussed in \cite{swap7,Daley}.
This method can be viewed either as diagonalizing the swap operator $\cal V$ to decompose it into a unitary operation plus detection \cite{swap7} or mapping the symmetric and antisymmetric states into the subspaces of even and odd numbered quanta in one mode \cite{Daley}.
{In this method, the unitary operation corresponds to 50:50 beam splitting whereas the detection corresponds to the parity measurement performed on one of the beam splitter outputs. Basically, the unitary operation maps the antisymmetric part of the initial state into states with odd number of excitations in the upper output port and the symmetric parts into states with even number of excitations. Because the symmetric and antisymmetric states are eigenstates of the SWAP operator with eigenvalues $1$ and $-1$, respectively, the parity measurement yields the expectation value of the SWAP operator.

The unitary operation is generated by an interaction Hamiltonian of the form 
\begin{equation}
H = -iJ(a_{k,a}^\dagger a_{k,b} - a_{k,b}^\dagger a_{k,a}), 
\label{bsham}
\end{equation}
where $k$ is the site index and subscripts $a$, $b$ denote copies of the density matrix. 50:50 beam splitting is obtained with $Jt = \pi/4$ and the parity measurement should be performed on the upper output port, i.e., the measurement operator is $(-1)^{n_{k,a}}$. 
The situation is depicted in Fig.~\ref{scheme_ho} for a single site case.
For $n-$partite states, the required detection is simply the parity of the total number of quanta in the upper modes of all sites, i.e., $(-1)^{\sum n_{k}}$, where $n_{k}$ denotes the photon number in the upper mode at site $k$.

\begin{figure}[t]
	\includegraphics[scale=0.5]{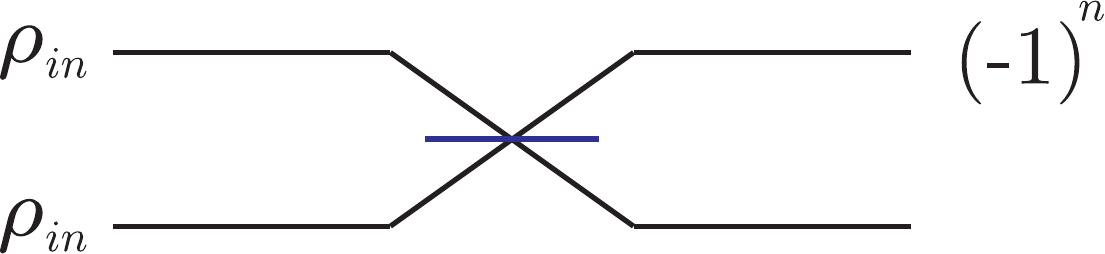}
	\caption{Direct purity measurement scheme for two harmonic oscillator states based on a beam splitter transformation and a parity measurement on one of the modes.}
	\label{scheme_ho}
\end{figure}

The parity measurement in general requires a photon number resolving detector or a full quantum state tomography (for non-Gaussian states).
The need of full tomography destroys the original intent of performing the purity measurement without full quantum state reconstruction (although the tomography is in a smaller subspace for multi-qubit states and therefore more manageable) and there is no photon number resolving detector to date.
This means that, using current technology, it would be difficult to measure $\mathcal{I}$ directly via this method for interesting (high-photon number) macroscopic harmonic oscillator states.
We note, however, that there are active ongoing researches in the development of photon number resolving detectors \cite{cat6, resolving2}.

\subsection{$n$-qubit states}

The generalization of the decomposition method for harmonic oscillator states to multi-qubit states is straightforward and yields the following result, which, as far as we are aware, have not been given elsewhere.
The beam-splitter-like unitary is generated by a Hamiltonian analogous to (\ref{bsham}):
	\begin{align}
	H_{BS} =
	J \left( \hat{\sigma}_{+} \otimes \hat{\sigma}_{-} + \hat{\sigma}_{-} \otimes \hat{\sigma}_{+} \right),
	\label{BSlike}
	\end{align}
performed for the duration $Jt = \pi/4$, acting on each pair of qubits.
Operators $\hat{\sigma}_{\pm}$ denote usual qubit raising and lowering operators.
We note that the Hamiltonian is proportional to $\hat{\sigma}_{x} \otimes \hat{\sigma}_{x} + \hat{\sigma}_{y} \otimes \hat{\sigma}_{y}$ and thus realizable in various spin systems. Due to the difference in commutation relationship between bosonic operators $a,a^\dagger$ and spin operators $\sigma_-,\sigma_+$, there is a slight change in the measurement part. Instead of the parity measurement on the first qubit, one has to measure the projection onto the state $|10\rangle$, i.e., the measurement operator is $(-1)^{\pjct{1}{1} \otimes \pjct{0}{0}}$.

There exists an alternative method which is more suitable for single-photon polarization qubits with
horizontally and vertically polarized single-photon states $|H\rangle$ and $|V\rangle$, for example.
In oder to see how this method works, note that the partial swap operator can be represented in the Bell-state basis as
\begin{equation}
{\cal V}_k =  P_{(k,+)} - P_{(k,-)} ,
\end{equation}
where  $P_{(k,+)} =\openone-P_{(k,-)}$ and $P_{(k,-)} =(|\zeta_-\rangle\langle\zeta_-|)_k$  with
 $|\zeta_{-}\rangle=(|H\rangle|V\rangle-|V\rangle|H\rangle)/\sqrt{2}$
denote the projectors onto the symmetric and antisymmetric subspaces of the $k$-th qubits, respectively.
Consider a detection scheme that gives a `click' at the $k$-th site when the qubits on the $k$-th site are in the antisymmetric  state $|\zeta_{-}\rangle$. Then we can decompose the swap operator into
	\begin{widetext}
	\begin{align}
	{\rm Tr} [{\cal V} \rho_a \otimes \rho_b]
	=& - \big( \textrm{The sum over probabilities of hearing clicks on
	an odd number of sites} \big) \nonumber \\ 
       &+ \big( \textrm{The sum over probabilities of all other click events} \big)
         	 \nonumber \\
	=&  1 - 2 \sum \big( \textrm{The probability of detecting odd number of events} \big)\equiv 1 - 2 \sum P_{\rm{odd}}.
	\label{nswap}
	\end{align}
	\end{widetext}
For single-photon polarization qubits, 
  the detection of the antisymmetric portion can be simply carried out by first mixing two photons in a beam splitter and detecting the coincidence events \cite{pc1,pc2,pc4}.
The same can be done for bosonic atoms, where the qubit is represented by two-species of atoms \cite{swap6}.
There, the beam splitter interaction is provided by the usual tight-binding form.

\section{Controlled-Swap schemes}
	\label{sect5}
	
In general, the difficulty in implementing a C-SWAP operation is in finding the right type of nonlinear interaction.
In this section, we summarize a previous proposal by Filip \cite{swap1} for harmonic oscillators for completeness and provide a new scheme for single-photon polarization qubits based on a general scheme proposed by Zhou \textit{et al.}~\cite{ZQ2011}.
We note that Lee \textit{et al.}'s method \cite{SML2013} to implement a C-SWAP gate can also be applied for our purpose here.

\subsection{Harmonic oscillator states}
In Ref.~\cite{swap1}, Filip noted that the C-SWAP gate on optical states can be written as
	\begin{align}
		U_{X} =
		U_{R}^{\dagger}\, U_{CPS}\, U_{R} ,
	\end{align}
where
	$U_{R}   = \exp \left[ (\pi/4)\, (\hat{a}_{0}^{\dagger} \hat{a}_{1} - \hat{a}_{1}^{\dagger} \hat{a}_{0}) \right]$
is the usual 50:50 beam splitter unitary,
	$U_{CPS} = \exp \left( i\pi\, \hat{a}_{1}^{\dagger} \hat{a}_{1} \pjct{V}{V} \right)$
is a controlled phase shift operator, and $\hat{a}_{0}$ ($\hat{a}_{1}$) refers to the annihilation operator for mode 0 (1).

To see that $U_{X}$ is indeed the required operator, we first note that $U_{R}$ mixes the annihilation operators as
	$U_{R}^{\dagger}\, \hat{a}_{0}\, U_{R} = \left( a_{1} + a_{0} \right)/{\sqrt{2}}$
and
	$U_{R}^{\dagger}\, \hat{a}_{1}\, U_{R} = \left( a_{1} - a_{0} \right)/{\sqrt{2}}$
so that
	\begin{equation}
		U_{X} = \exp \left[ i\, \frac{\pi}{2}\, \big( \hat{a}_{1} - \hat{a}_{0} \big)^{\dagger}\, \big( \hat{a}_{1} - \hat{a}_{0} \big)\, \pjct{V}{V}\, \right] .
	\end{equation}
From this expression it is easily seen that
	\begin{align}
		U_{X}\, \hat{a}_{0}\, \pjct{V}{V}\, U_{X}^{\dagger} &= \hat{a}_{1}\, \pjct{V}{V}, \nonumber \\
		U_{X}\, \hat{a}_{1}\, \pjct{V}{V}\, U_{X}^{\dagger} &= \hat{a}_{0}\, \pjct{V}{V}, \nonumber \\
		U_{X}\, \hat{a}_{0}\, \pjct{H}{H}\, U_{X}^{\dagger} &= \hat{a}_{0}\, \pjct{H}{H}, \nonumber \\
		U_{X}\, \hat{a}_{1}\, \pjct{H}{H}\, U_{X}^{\dagger} &= \hat{a}_{1}\, \pjct{H}{H} ,
	\end{align}
completing the proof that the unitary operator $U_{X}$ is the sought-after C-SWAP operator.

\begin{figure*}[t]
	\includegraphics[scale=0.18]{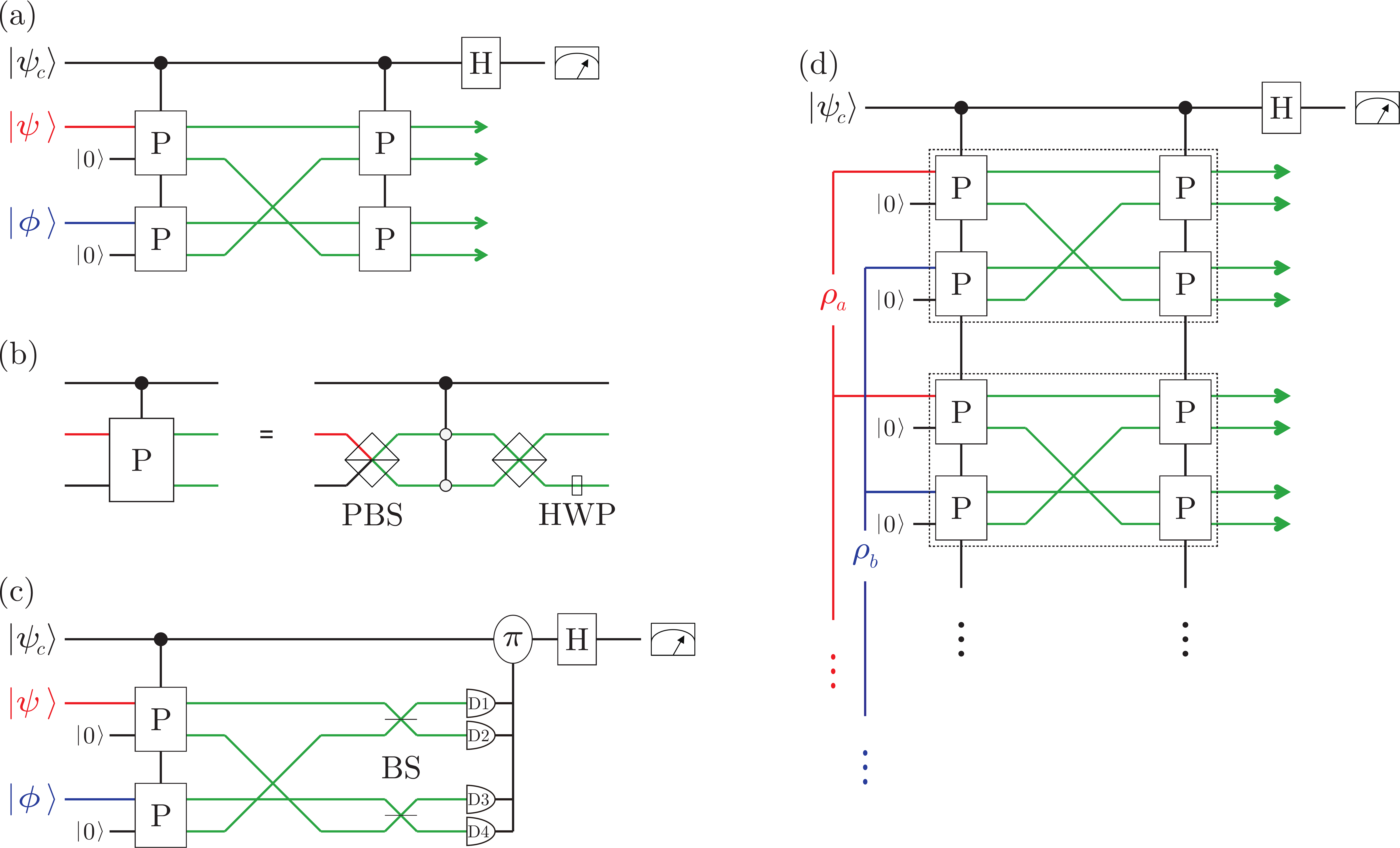}
		\caption{(Color online) (a) C-SWAP gate for two single-photon polarization qubits. P represents the conditional change of the paths as a part of a CP gate, and H represents a Hadamard gate.
				(b) Construction of a CP gate. PBS represents a polarization beam splitter and HWP a half-wave plate.
				(c) Alternative way to construct a C-SWAP gate for two single-photon polarization qubits. BS represents a 50:50 beam splitter, $\pi$ represents a $\pi$ phase shifter, and D1 to D4 are photodetectors.
				(d) C-SWAP gate for two multi-photon polarization qubits.}
	\label{fig-optical}
\end{figure*}

The form of the controlled phase shift operator arises often in the discussion of quantum nondemolition  measurements \cite{qnd}, which have been discussed in the settings of cavity quantum electrodynamics (QED) \cite{Brune,Nogues}, trapped ions \cite{Filho, Davidovich}, and electromagnetically induced transparency (EIT) \cite{Imoto, Greentree} among others.
The cavity QED scheme involves dispersive atom-field coupling, giving rise to an effective interaction of the form
	$\hat{a}^{\dagger} \hat{a}\,\hat{\sigma}_{+} \hat{\sigma}_{-}$ \cite{Nogues}.
The trapped ion version is based on a realization of the Hamiltonian
	$H = \lambda\, \hat{a}^{\dagger}\hat{a}\, \left( \hat{\sigma}_{+} + \hat{\sigma}_{-} \right)$, 
which arises when the qubit is driven resonantly with a small Rabi frequency (smaller than the vibrational energy)
{\cite{Gerry}}.
The EIT version utilizes the cross-Kerr nonlinearity
	$\hat{a}^{\dagger}\hat{a}\,\hat{b}^{\dagger}\hat{b}$,
i.e., the role of the qubit is now played by a single photon thus requiring large cross phase modulation \cite{Imoto}.

\subsection{$n$-qubit states}

We now discuss a feasible scheme to detect macroscopic quantumness $\mathcal{I}$ for multi-qubit states based on the linear optical implementation of controlled-unitary gates introduced in Ref.~\cite{ZQ2011}. Here, polarization states of photons are used to encode the qubits.

Let us start by describing the simplest case of measuring the overlap between two qubit states $\ket{\psi}$ and $\ket{\phi}$, as depicted in Fig.~\ref{fig-optical}(a).
The basic idea is to add auxiliary modes, conditionally interchange the original and auxiliary modes, perform swap on the auxiliary modes only, and interchange the paths conditionally once again.
The main primitive used in this method is the controlled-path (CP) gate, which interchanges the paths of the two incoming photons only if the control qubit is in, say the state $\ket{V}$.

To see how the scheme works, consider the initial state of the whole system, given the control qubit state
	$\ket{\psi_{c}} = c_{h}\ket{H} + c_{v}\ket{V}$:
\begin{equation}
		c_{h}\ket{H}\ket{\psi}\ket{0}\ket{\phi}\ket{0} + c_{v}\ket{V}\ket{\psi}\ket{0}\ket{\phi}\ket{0} .
	\label{cse1}
\end{equation}
The first two CP gates are applied as depicted Fig.~\ref{fig-optical}(a).
As shown in Fig.~\ref{fig-optical}(b), a CP gate can be implemented using two polarization beam splitters (PBSs), two C-NOT gates and a half-wave plate (HWP) \cite{ZQ2011}.
If the control qubit $\ket{\psi_{c}}$ is in $\ket{H}$, CP does nothing.
If it is in $\ket{V}$, however, CP changes the paths of $\ket{\psi}$ and $\ket{0}$ (and  $\ket{\phi}$ and $\ket{0}$).
Therefore, the state after the first two CP gates becomes
\begin{equation}
		c_{h}\ket{H}\ket{\psi}\ket{0}\ket{\phi}\ket{0} + c_{v}\ket{V}\ket{0}\ket{\psi}\ket{0}\ket{\phi} .
	\label{cse2}
\end{equation}
We then change the paths of the third and the fifth qubits as shown in Fig.~\ref{fig-optical}(a) to obtain 
\begin{equation}
		c_{h}\ket{H}\ket{\psi}\ket{0}\ket{\phi}\ket{0} + c_{v}\ket{V}\ket{0}\ket{\phi}\ket{0}\ket{\psi} .
	\label{cse3}
\end{equation}
The next two CP gates reverse the initial controlled-path change, so that the final state becomes
\begin{equation}
		c_{h}\ket{H}\ket{\psi}\ket{0}\ket{\phi}\ket{0} + c_{v}\ket{V}\ket{\phi}\ket{0}\ket{\psi}\ket{0} .
	\label{cse4}
\end{equation}

Figure~\ref{fig-optical}(c) presents an alternative way to construct an effective C-SWAP gate in the circuit.
In place of the two CP gates after the path exchange, two beam splitters and four detectors are used.
Furthermore, depending on the measurement outcomes of the detectors, a $\pi$ phase shift may be performed on the target qubit. 
The state after the first CP gates and the path exchange is in Eq.~(\ref{cse3}).
As states
$\ket{\psi}$ and $\ket{\phi}$ are single photon states, there are four possibilities of detection after the two 50:50 beam splitters: (D1, D3), (D1, D4), (D2, D3), (D2, D4) where the detector numbers are indicated in Fig.~\ref{fig-optical}(c).
The final state that gives rise to these possibilities are
	\begin{align}
		c_{h}\ket{H}\ket{\psi}\ket{0}\ket{\phi}\ket{0} + c_{v}\ket{V}\ket{\phi}\ket{0}\ket{\psi}\ket{0}, \\
		- c_{h}\ket{H}\ket{\psi}\ket{0}\ket{0}\ket{\phi} + c_{v}\ket{V}\ket{\phi}\ket{0}\ket{0}\ket{\psi} ,\\
		- c_{h}\ket{H}\ket{0}\ket{\psi}\ket{\phi}\ket{0} + c_{v}\ket{V}\ket{0}\ket{\phi}\ket{\psi}\ket{0} ,\\
		~c_{h}\ket{H}\ket{0}\ket{\psi}\ket{0}\ket{\phi} + c_{v}\ket{V}\ket{0}\ket{\phi}\ket{0}\ket{\psi},
	\end{align}
	in the same order.
For the (D1, D4) or (D2, D3) case, a $\pi$ phase shift should be performed on the target qubit.
In this way, the same process for the overlap measurement can be performed using only half the number of C-NOT gates compared to the scheme in Fig.~\ref{fig-optical}(a).
It is interesting to note that in the expense of the reduced number of C-NOT gates, exactly the same number of photodetectors have been inserted.

The method presented in Fig.~\ref{fig-optical}(c) may be useful because an all-optical C-NOT gate requires a pre-arranged entangled state and is typically nondeterministic \cite{cnot1, cnot2}. 
The whole process described in this subsection can be extended to any two states with an arbitrary number of photons in each state.
For example, an extension of the scheme in Fig.~\ref{fig-optical}(a) with two states $\rho_{a}$ and $\rho_{b}$ is depicted in Fig.~\ref{fig-optical}(d).
In the same manner, the schemes in Fig.~\ref{fig-optical}(c) can be extended to any two states with an arbitrary number of photons.

\section{Experimental imperfections}
	\label{sect67}
So far, we have discussed various overlap measurement schemes, any of which can be adopted to detect  macroscopic quantumness $\mathcal{I}$ as detailed in the previous sections. In this section, we discuss the effects of two experimental imperfections: finite time resolution and imperfect detection efficiencies.

\subsection{Effects of coarse-grained measurement}
	\label{sect6}
In an experiment, one is forced to use a finite value of $\Delta \tau$ in Eq.~(\ref{coarse-grained MQ}), which leads to a certain amount of error in detecting $\mathcal{I}$. Below, we discuss through specific examples the effects of such coarse-grained implementations. The first example is SCSs \cite{cat1,Ourjoumtsev,cat4,cat5,cat6} under the decoherence effect described by Eq.~(\ref{dmodel}): 
	\begin{align}
		\rho_{\rm{scs}}(\tau) = 
		\mathcal{N}\; \Big\{
		&\pjct{t\alpha}{t\alpha} + \pjct{-t\alpha}{-t\alpha} \nonumber\\
		&+\Gamma(\tau) \left( \pjct{t\alpha}{-t\alpha} + \pjct{-t\alpha}{t\alpha} \right) \Big\} ,
	\label{SCSunderDecoherence}
	\end{align}
where $t = \exp(-\tau/2)$, $\Gamma(\tau) = \exp \left[-2\; (1-e^{-\tau})\; \alpha^2 \right]$, and $\mathcal{N}$ is the normalization factor.
Macroscopic quantumness is obtained as
	\begin{equation}
		\mathcal{I}(\alpha,\tau) = 
		\langle\hat{n}(0)\rangle\, \frac{e^{-\tau}\, \sinh{\left[\, 2\, (2\, e^{-\tau} - 1 )\, \alpha^{2} \right]}}{\sinh{2\, \alpha^{2}}} ,
\end{equation}
where $\langle\hat{n}(0)\rangle = \alpha^{2}\, \tanh{\alpha^{2}}$ denotes  the average number of photons at $\tau=0$. This type of decoherence can be implemented using a beam splitter of reflectivity $r=\sqrt{1-e^{-\tau}}$. Note that for a pure SCS, the measure $\mathcal{I}$ equals the average photon number, i.e., $\mathcal{I}(\alpha,0) = \langle\hat{n}(0)\rangle$.

Now, suppose that one intends to measure $\mathcal{I}$ for $\rho_{\rm{scs}}(\tau)$. This requires the purity of the state be measured twice at $\tau$ and $\tau+\Delta \tau$. As explained earlier, one may use two beam splitters of appropriate ratios to obtain $\rho_{\rm{scs}}(\tau)$ and $\rho_{\rm{scs}}(\tau+\Delta \tau)$. 
The purity of the SCS is
	\begin{align}
		&\mathcal{P}(\rho_{scs}(\tau))=\Tr[\rho_{scs}^{2}(\tau)] \nonumber\\
		&=\tanh{\alpha^{2}}\frac{2+2\cosh{2\,\alpha^{2}}+\cosh{\left[\, 2\, (2\, e^{-\tau} - 1 )\, \alpha^{2} \right]}}{2 \sinh{2\,\alpha^{2}}}.
		\label{eq:purityscs}
	\end{align}
Using Eqs.~(\ref{coarse-grained MQ}) and (\ref{eq:purityscs}), we obtain and plot the measured quantity, $\mathcal{I}_{\Delta\tau}(\rho_{\rm{scs}})$, as a function of $\Delta\tau$ for several choices of $\tau$ in Figs.~\ref{fig7abcd}(a) and \ref{fig7abcd}(b). 
Of course, the measured quantity $\mathcal{I}_{\Delta\tau}(\rho_{\rm{scs}})$ approaches the precise quantity $\mathcal{I}(\alpha,\tau)$ when $\Delta\tau\rightarrow 0$.
One can see that a smaller value of $\Delta\tau$ is required for larger $\alpha$ in order to precisely assess the value of $\mathcal{I}$.

\begin{figure}[t]
	\includegraphics[scale=0.4]{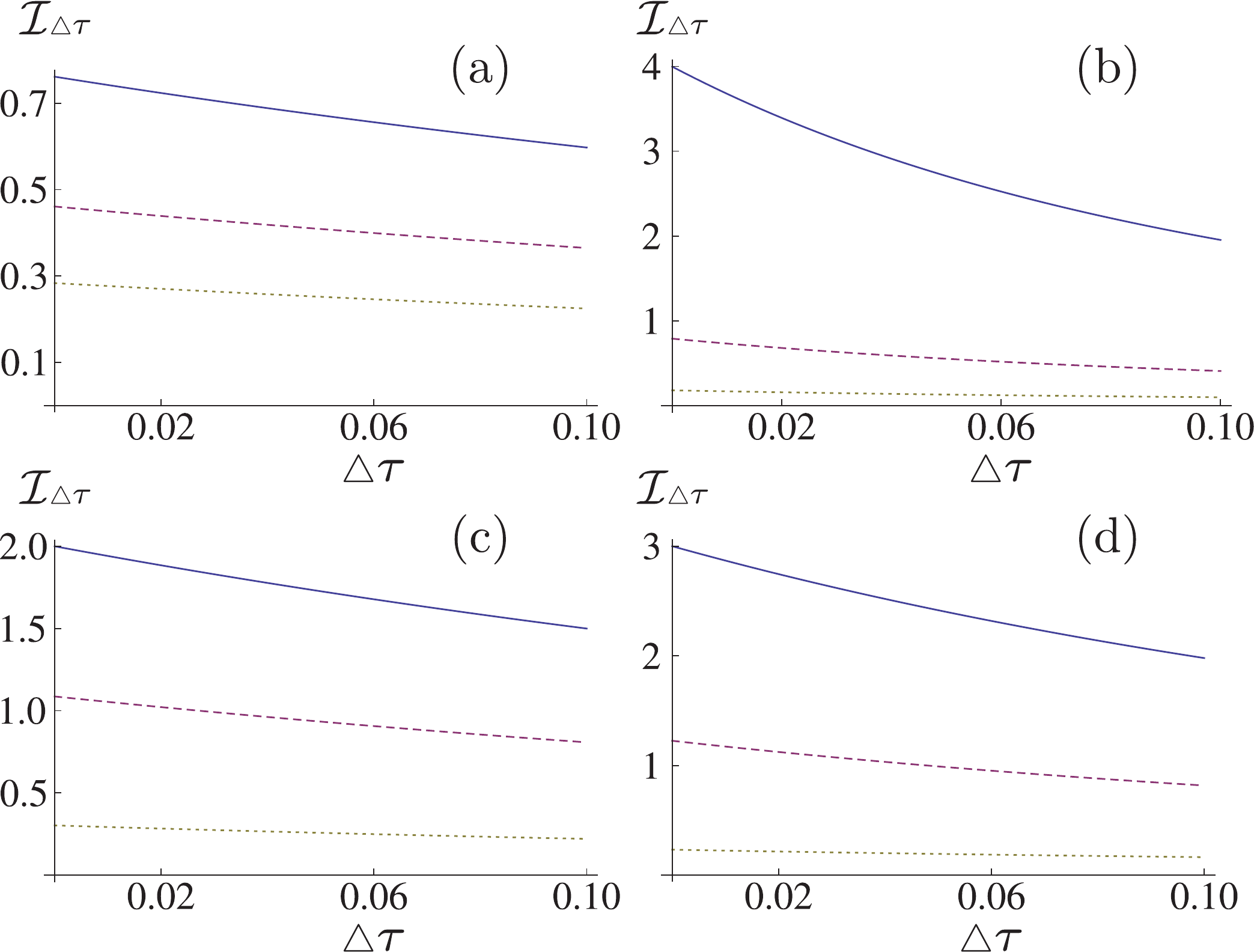}
	\caption{(Color online) Coarse-grained version of macroscopic quantumness, $\mathcal{I}_{\Delta\tau}$, as a function of $\Delta\tau$ that can be measured via the proposed scheme.
			The values are obtained for SCSs under decoherence $\rho_{\rm{scs}}(\tau)$ with (a) $\alpha=1$ and (b) $\alpha=2$ as well as number states (c) $\ket{2}$ and (d) $\ket{3}$ under decoherence.
			The solid, dashed and dotted curves in each figure corresponds to $\tau=0$, $\tau=0.1$ and $\tau=0.2$, respectively.}
	\label{fig7abcd}
\end{figure}

The next example we consider are photon number states $\ket{2}$ and $\ket{3}$. These states evolve under the decoherence as
	\begin{align}
		\rho_{\ket{2}}(\tau) 
		=&e^{-2\tau}\pjct{2}{2}+2\,e^{-\tau}{\cal G}\pjct{1}{1} +{\cal G}^{2}\pjct{0}{0} , \\
		\rho_{\ket{3}}(\tau)
		=&e^{-3\tau}\pjct{3}{3}+3\,e^{-2\tau}{\cal G}\pjct{2}{2} \nonumber\\
		&+3\,e^{-\tau}{\cal G}^{2}\pjct{1}{1}+{\cal G}^{3}\pjct{0}{0} ,
	\end{align}
where ${\cal G}=1-e^{-\tau}$, and their purities are  obtained as
	\begin{align}
		\mathcal{P}(\rho_{\ket{2}}(\tau)) 
		&=e^{-4 \tau} + 4\,e^{-2 \tau} \mathcal{G}^2 + \mathcal{G}^4,  \label{eq:p2} \\
		\mathcal{P}(\rho_{\ket{3}}(\tau))
		&=e^{-6 \tau} + 9\,e^{-4 \tau} \mathcal{G}^2 + 9\,e^{-2 \tau} \mathcal{G}^4 + \mathcal{G}^6.  \label{eq:p3}
		\end{align}
The precise values of macroscopic quantumness for $\rho_{|2\rangle}(\tau)$ and $\rho_{|3\rangle}(\tau)$ are then
	\begin{align}
		\mathcal{I}(\ket{2},\tau) =&2\,e^{-4 \tau} - 4\,e^{-3 \tau} \mathcal{G} + 4\,e^{-2 \tau} \mathcal{G}^{2} - 2\,e^{-\tau} \mathcal{G}^3, \\
		\mathcal{I}(\ket{3},\tau) =&3\,e^{-6 \tau} - 9\,e^{-5 \tau} \mathcal{G} + 18\,e^{-4 \tau} \mathcal{G}^{2} \nonumber\\	
		& - 18\,e^{-3 \tau} \mathcal{G}^3 + 9\,e^{-2 \tau} \mathcal{G}^{4} - 3\,e^{-\tau} \mathcal{G}^{5}.
	\end{align}
respectively. 
Figures~\ref{fig7abcd}(c) and \ref{fig7abcd}(d) plot the coarse-grained measure, $\mathcal{I}_{\Delta\tau}$, as a function of $\Delta\tau$. 
We observe a similar behavior to that of the SCSs;
a smaller value of $\Delta\tau$ is required for a larger number of photons in order to precisely represent $\mathcal{I}$.

We conclude that given a small enough value of $\Delta\tau$,
the coarse-grained measure $\mathcal{I}_{\Delta\tau}$ is not too different from $\mathcal{I}$, and one can safely propose $\mathcal{I}_{\Delta\tau}$ itself as the measure of macroscopic quantumness.
The latter is then a coarse-grained version of $\mathcal{I}$: 
	\begin{align}
		\frac{1}{\Delta\tau}\int_{0}^{\Delta\tau} \ud t\, \mathcal{I}(\rho) = \mathcal{I}_{\Delta\tau}(\rho) ,
	\end{align}
which follows from Eqs.~(\ref{eq5}) and (\ref{coarse-grained MQ}).

\subsection{Effects of detection inefficiency}
	\label{sect7}
Another imperfection present in experiments is detection inefficiency, which we investigate here for both the direct and C-SWAP schemes.

\subsubsection{Direct measurement scheme}
We first give our attention to harmonic oscillator states
where experimentally demanding parity measurements are required.
A single mode optical state $\rho_{in}$ (here, the subscription $in$ means `input') can be expressed in the Glauber-Sudarshan representation
	\begin{align}
		\rho_{in} = 
		\int \ud^{2}\alpha\; P_{in}(\alpha) \pjct{\alpha}{\alpha} ,
	\end{align}
where $P(\alpha)$ is called the $P$ function.
Using this representation, we can calculate how the input states evolve under the direct measurement scheme described in Sec.~\ref{sect4a} (see Fig.~\ref{scheme_ho}).
The state in the upper arm after a 50:50 beam splitter, $\rho_{mid}$, is  
	\begin{align}
		\rho_{mid} &= 
		\Tr \left[\, U_{R}\, \big(\,\rho_{in}\otimes\rho_{in}\, \big)\, U_{R}^{\dagger} \,\right] \nonumber\\
		&= \int \ud^{2}\alpha\; \ud^{2}\beta \; P_{in}(\alpha) P_{in}(\beta) \, \pjct{\frac{\alpha+\beta}{\sqrt{2}}}{\frac{\alpha+\beta}{\sqrt{2}}} ,
	\end{align}
where the trace is taken over the field in the bottom arm.
An imperfect detector with efficiency $\eta$ is modeled by placing a beam splitter with transmittance $\eta$ in front of a perfect detector.
The beam splitter  operator is
	$U_{\eta} = \exp \left[\, \theta\, (\, \hat{a}^{\dagger}_{0} \hat{a}_{1} - \hat{a}_{1} \hat{a}^{\dagger}_{0}\, )\, \right]$
with  transmittance $\eta = \cos^{2}{\theta}$. When $U_{\eta}$ is applied to two coherent states, it yields $U_{\eta}\ket{\alpha}\ket{\beta}=\ket{\sqrt{\eta}\,\alpha+\sqrt{1-\eta}\,\beta}\ket{-\sqrt{1-\eta}\,\alpha+\sqrt{\eta}\,\beta}$.
After passing through such a beam splitter, the state $\rho_{out}$ to be detected is
	\begin{align}
		&\rho_{out} =
		\Tr \left[\, U_{\eta}\, \big(\, \rho_{mid} \otimes \pjct{0}{0}\, \big)\, U_{\eta}^{\dagger}\, \right] \nonumber\\
		&= \int \ud^{2}\alpha\; \ud^{2}\beta \; P_{in}(\alpha) P_{in}(\beta) \, \pjct{\sqrt{\frac{\eta}{2}}(\alpha+\beta)}{\sqrt{\frac{\eta}{2}}(\alpha+\beta)} ,
	\label{imdetout}
	\end{align}
where the trace is now over the reflected arm of the beam splitter.
The final result of the measured purity  $\widetilde{\mathcal{P}}$  is then given by
	\begin{align}
		\widetilde{\mathcal{P}}
		&=\sum_{n} (-1)^{n} \bra{n}\, \rho_{out}\, \ket{n} \nonumber\\		
		&=\int\!\!\!\int \ud^{2} \alpha\, \ud^{2} \beta\; P_{in}(\alpha)\, P_{in}(\beta)\, e^{-\eta\, (\alpha+\beta)^{2}} ,
		\label{eq:ptil}
	\end{align}
which is obtained using 
$\sum_{n} (-1)^{n} \langle n \pjct{\alpha}{\alpha} n \rangle 
= e^{-2\alpha^{2}}$.

\begin{figure}[t]
\centering
	\includegraphics[scale=0.40]{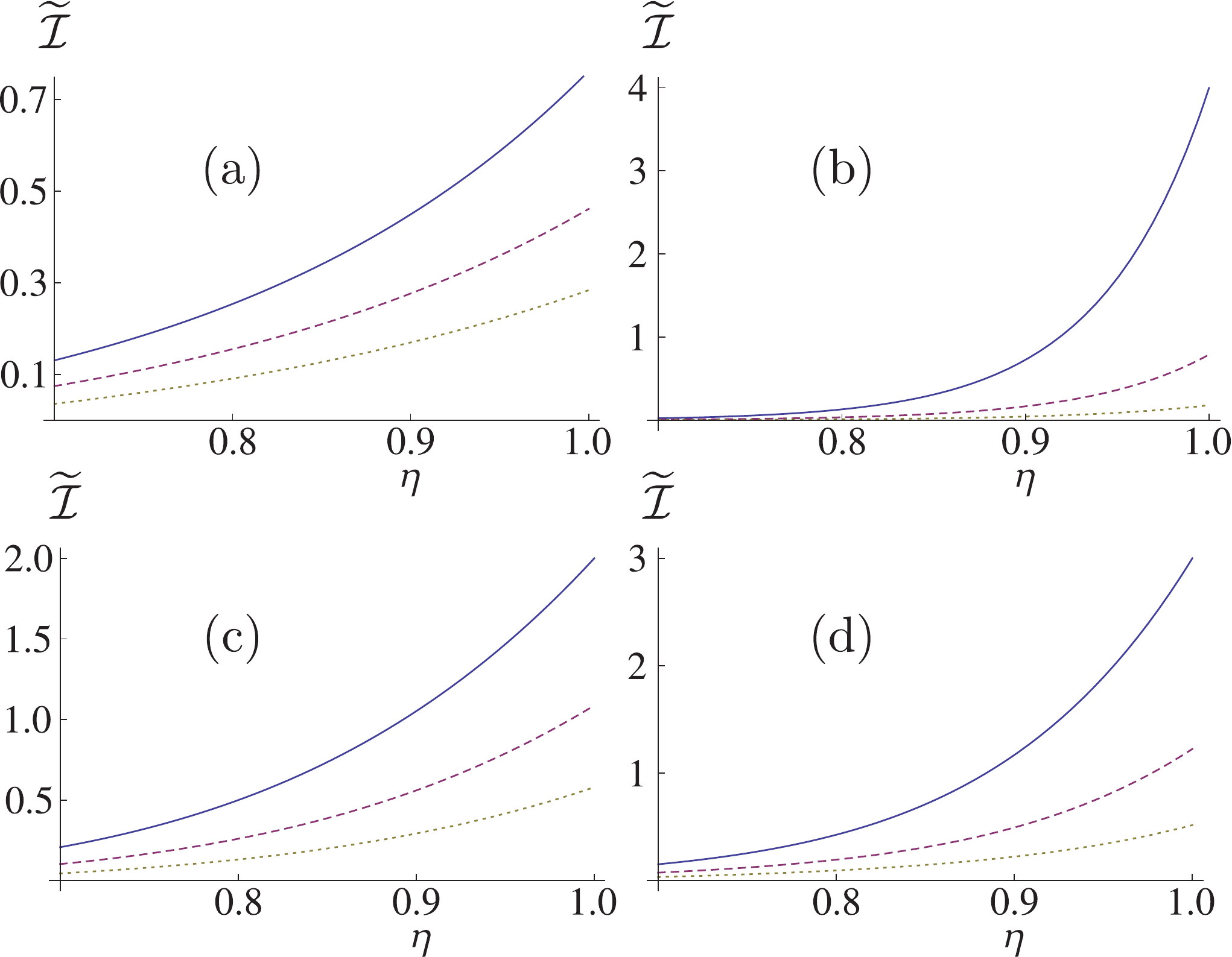}
	\caption{(Color online) Measured macroscopic quantumness $\widetilde{\mathcal{I}}$ for SCSs with (a) $\alpha=1$ and (b) $\alpha=2$ as well as number states (c) $|2\rangle$ and (d) $|3\rangle$ as a function of detection efficiency $\eta$.
			The solid, dashed and dotted curves correspond to $\tau=0$, $\tau=0.1$ and $\tau=0.2$, respectively, in each figure.}
\label{SCS}
\end{figure}

Using this formula we will investigate the effects of detection inefficiency through the two examples used above.
The first example is an SCS under the decoherence effect described by Eq.~(\ref{SCSunderDecoherence}).
Its measured purity with detection efficiency $\eta$ is
	\begin{align}
		\widetilde{\mathcal{P}} = 
		\tanh{\alpha^{2}}\frac{ 2 + \cosh{2\, \alpha^{2}} + \cosh{\big[ 2\, (2\, \eta\, e^{-\tau} -1)\, \alpha^{2} \big]} }{2\, \sinh{2\, \alpha^{2}}},
	\end{align}
which yields the measured macroscopic quantumness
	\begin{align}
		\widetilde{\mathcal{I}}(\alpha,\eta,\tau)
		&=-\frac{1}{2}\, \frac{\ud\,\widetilde{\mathcal{P}}}{\ud\,\tau} \nonumber\\
		&=\alpha^2\, \tanh{\alpha^2}\, \frac{\eta\, e^{-\tau}\, \sinh{\left[ 2\, (2\, \eta\, e^{-\tau} - 1)\, \alpha^2 \right]}}{\sinh{2\,\alpha^2}} .
	\end{align}
$\widetilde{\mathcal{I}}(\alpha,\eta,\tau)$ as a function of $\eta$ for three $\tau$ cases is plotted in Figs.~\ref{SCS}(a) and \ref{SCS}(b).

The measured macroscopic quantumness for the number states are:
	\begin{align}
		\widetilde{\mathcal{I}} \left( \ket{2},\eta,\tau \right) = 
		&-2\,\eta\,\mathrm{e}^{-\tau} +10\,\eta^{2}\,\mathrm{e}^{-2\tau} \nonumber \\
		&-18\,\eta^{3}\,\mathrm{e}^{-3\tau} +12\,\eta^{4}\,\mathrm{e}^{-4\tau},
	\end{align}
	\begin{align}
		\widetilde{\mathcal{I}} \left( \ket{3},\eta,\tau \right) =
		&-3\,\eta\,\mathrm{e}^{-\tau} +24\,\eta^{2}\,\mathrm{e}^{-2\tau} -84\,\eta^{3}\,\mathrm{e}^{-3\tau} \nonumber \\
		&+156\,\eta^{4}\,\mathrm{e}^{-4\tau} -150\,\eta^{5}\,\mathrm{e}^{-5\tau} +60\,\eta^{6}\,\mathrm{e}^{-6\tau},		
	\end{align}
which are plotted in Figs.~\ref{SCS}(c) and \ref{SCS}(d). 
In both the cases, the measured values degrade as $\eta$ decreases, while the number states are slightly more robust against the detection inefficiency.

We also briefly address inefficiency in the direct measurement schemes for qubits in Sec. 4.B. 
Consider the first method involving the spin-beam-splitter operation described by the Hamiltonian in Eq.~(\ref{BSlike}) and a measurement of  $(-1)^{\pjct{1}{1} \otimes \pjct{0}{0}}$. Assume that there are detectors that can distinguish between $|0\rangle$ and $|1\rangle$, whose inefficiency is such that these two states can be missed with equal probability. Then the detector inefficiency has no effect on the value of the measured observable, because it simply changes the total number of counts that has no effect on the measurement probabilities. 
Similarly, the effect of known detector inefficiency in the second method summarized by Eq.~(\ref{nswap}) can be taken into account in the calculation of the overlap from raw data. Here, the inefficiency in measuring the antisymmetric part of a subsystem in single site (involving two qubits) can be determined prior to the overlap measurement and subsequently used to counterbalance the bias due to the detector inefficiency.

\subsubsection{Controlled-Swap Schemes}
In the C-SWAP schemes, the control qubit (e.g. a polarized photon) may be lost before the detector because of the detection inefficiency. The result is a reduced number of counts, but
since the detection efficiency is usually known, the correct number of counts can always be inferred, yielding straightforwardly the correct value of the purity. Furthermore, because there is only one qubit that needs to be detected at the final detector unlike in the direct measurement schemes, the effect of detection inefficiency can be easily corrected.
This is an advantage of the C-SWAP schemes compared to the direct measurement schemes.

The correction of the inefficiency in detectors D1 to D4 in Fig.~\ref{fig-optical}(c) is even easier, because the runs that do not have two detector clicks are simply discarded.
On the other hand, if the fidelity of the CP gates is limited, it may cause the outcomes at D1-D4 to be biased and therefore degrade the accuracy of the final result. Such effects may be caused by the limited fidelity of the resource entangled states or mode mismatching at beam splitters.

\section{Summary and discussions}
	\label{sect8}
In this paper, we have briefly reviewed macroscopic quantumness $\mathcal{I}$ and described experimental methods to measure the quantity without requiring quantum state tomography, both for harmonic oscillator and qubit systems. 
Broadly, these methods can be sorted into two categories: Direct measurement schemes and C-SWAP schemes. After summarizing previously proposed schemes to achieve overlap measurements that can be applied to our purpose, we proposed a new C-SWAP scheme for single-photon polarization qubits. We have investigated effects of experimental limitations such as imperfect detection efficiencies and finite time resolutions. 

Our method provides means to detect the macroscopic quantumness of various states such as SCSs \cite{cat1,Ourjoumtsev,cat4,cat5,cat6} in experiments without the need of quantum state tomography. For fully tomography, the number of measurement settings increases exponentially with the size of the state under consideration, whilst our scheme requires only two measurement settings, i.e., {\it with and without} induced decoherence.
For qubit tomography, the number of measurement settings increases as $\sim 4^{n}$ where $n$ is the number of qubits \cite{tomo1, tomo2}.
In contrast, only two settings are required for our scheme where simply the number of required detectors linearly increases as $\sim n$ as shown in Fig.~\ref{npartite}.
A similar argument is applied to the case of harmonic oscillator states where many homodyne settings are necessary to perform a full tomography of an arbitrary state \cite{tomo3}, while our method only requires two settings (again, with and without induced decoherence) with parity measurements.

Since the measure can be applied to mixed states \cite{LJ2011}, highly mixed states with nonclassical features \cite{mix1,mix1-2,mix2,Nha2008} can also be analyzed.
Furthermore, macroscopic superpositions and entanglement of spin systems \cite{spincat} under dephasing effects may be simulated using the proposed optical setup to experimentally explore their behavior in terms of macroscopic quantumness.

\acknowledgements
C.N. would like to thank Su-Yong Lee for helpful discussions.
H.J. and S.B. were supported by the National Research Foundation of Korea (NRF) grant funded by the Korean Government (MSIP) (No. 3348-20100018).
C.N. and D.G.A. are supported by the Singapore National Research Foundation and Ministry of Education (partly through the Academic Research Fund Tier 3 MOE2012-T3-1-009).
T.C.R. was supported by the Australian Research Council Centre of Excellence for Quantum Computation and Communication Technology (Project number CE11000102).


\begin{thebibliography}{99}

\bibitem{Schrodinger}
E.~Schr\"{o}dinger,
``Die gegenw\"{a}rtige Situation in der Quantenmechanik'',
Die Naturwissenschaften \textbf{23}, 807 -812 (1935).

\bibitem{MonroeCat}
C.~Monroe, D.~M.~Meekhof, B.~E.~King and D.~J.~Wineland,
``A ``Schr\"{o}dinger Cat'' Superposition State of an Atom'',
Science \textbf{272}, 1131-1136 (1996). 

\bibitem{C60}
M.~Arndt, O.~Nairz, J.~Vos-Andreae, C.~Keller, G.~van~der~Zouw and A.~Zeilinger,
``Wave-particle duality of $C_{60}$ molecules'',
Nature \textbf{401}, 680-682 (1999). 

\bibitem{SQUID2}
J.~R.~Friedman, V.~Patel, W.~Chen, S.~K.~Tolpygo and J.~E.~Lukens,
``Quantum superposition of distinct macroscopic states'',
Nature \textbf{406}, 43-46 (2000).

\bibitem{SQUID1}
C.~H.~van~der~Wal, A.~C.~J.~ter~Haar, F.~K.~Wilhelm, R.~N.~Schouten, C.~J.~P.~M.~Harmans, T.~P.~Orlando, S.~Lloyd and J.~E.~Mooij,
``Quantum superposition of Macroscopic Persistent-Current States'',
Science \textbf{290}, 773-777 (2000).

\bibitem{Ourjoumtsev}
A.~Ourjoumtsev, H.~Jeong, R.~Tualle-Brouri, and P.~Grangier,
``Genertation of optical `Schr\"{o}dinger cats' from photon number states'',
Nature \textbf{448}, 784-786 (2007).

\bibitem{Gao}
W.-B.~Gao, C.-Y.~Lu, X.-C.~Yao, P.~Xu, O.~G\"{u}hne, A.~Goebel, Y.-A.~Chen, C.-Z.~Peng, Z.-B.~Chen and J.-W.~Pan,
``Experimental demonstration of a hyper-entangled ten-qubit Schr\"{o}dinger cat state'',
Nature Physics \textbf{6}, 331-335 (2010).

\bibitem{Afek}
I.~Afek, O.~Ambar, and Y.~Silberberg,
``High-NOON States by Mixing Quantum and Classical Light'',
Science \textbf{328}, 879-881 (2010).

\bibitem{Leggett}
A.~J.~Leggett,
Macroscopic Quantum Systems and the Quantum Theory of Measurement'',
Prog. Theor. Phys. Suppl. \textbf{69}, 80-100 (1980);
``Testing the limits of quantum mechanics: motivation, state of play, prospects'',
J. Phys.: Condens. Matter \textbf{14}, R415 (2002).

\bibitem{Dur}
W.~D\"{u}r, C.~Simon, and J.~I.~Cirac,
``Effective Size of Certain Macroscopic Quantum Superpostions'',
Phys. Rev. Lett. \textbf{89}, 210402 (2002).

\bibitem{Shimizu2002}
A.~Shimizu and T.~Miyadera,
``Stability of Quantum States of Finite Macroscopic Systems against Classical Noises, Perturbations from Environments, and Local Measurements'',
 Phys. Rev. Lett. \textbf{89} 270403 (2002).

\bibitem{Bjork}
G.~Bj\"{o}rk and P.~G.~L.~Mana,
``A size criterion for macroscopic superposition states'',
J. Opt. B \textbf{6}, 429-436 (2004).

\bibitem{Shimizu2005}
A.~Shimizu and T.~Morimae,
``Detection of Macroscopic Entanglement by Correlation of Local Observables'',
Phys. Rev. Lett. \textbf{95} 090401 (2005).

\bibitem{Cavalcanti}
E.~G.~Cavalcanti and M.~D.~Reid,
``Signatures for Generalized Macroscopic Superpositions'',
Phys. Rev. Lett. \textbf{97}, 170405 (2006).

\bibitem{Korsbakken}
J.~I.~Korsbakken, K.~B.~Whaley, J.~Dubois, and J.~I.~Cirac,
``Measurement-based measure of the size of macroscopic quantum superpositions'',
Phys. Rev. A \textbf{75}, 042106 (2007).

\bibitem{Mar}
F.~Marquardt, B.~Abel, and J.~von~Delft,
``Measuring the size of a quantum superposition of many-body states'',
Phys. Rev. A \textbf{78}, 012109 (2008).

\bibitem{Korsbakken2}
J.~I.~Korsbakken, F.~K.~Wilhelm, and K.~B.~Whaleyl,
``The size of macroscopic superposition states in flux qubits'',
Europhysics Letters \textbf{89}, 30003  (2010).

\bibitem{LJ2011}
C.-W.~Lee and H.~Jeong,
``Quantification of Macroscopic Quantum Superpositions within Phase Space'',
Phys. Rev. Lett. \textbf{106}, 220401 (2011).

\bibitem{Flowis2012}
F.~Fr\"{o}wis and W.~D\"{u}r,
``Measures of macroscopicity for quantum spin systems'',
New J. Phys. \textbf{14}, 093039 (2012).

\bibitem{Nim2013}
S.~Nimmrichter and K.~Hornberger,
``Macroscopicity of Mechanical Quantum Superposition States'',
Phys.~Rev.~Lett. \textbf{110}, 160403 (2013).

\bibitem{Sek2014}
P.~Sekatski, N.~Sangouard, and N.~Gisin,
``Size of qunatum superspositions as measured with classcial detectors'',
Phys.~Rev.~A \textbf{89}, 012116 (2014).

\bibitem{Sek2014b} 
P.~Sekatski, N.~Gisin, and N.~Sangouard,
``How difficult it is to prove the quantumness of macroscopic states?'',
Phys.~Rev.~Lett. \textbf{113}, 090403 (2014).

\bibitem{Gir2014}
D.~Girolami,
``Observable measure of quantum coherence in finite dimensinal systems'',
http://arxiv.org/abs/1403.2446.

\bibitem{HJRev} 
H.~Jeong, M.~Kang and H.~Kwon
``Characterizations and quantifications of macroscopic quantumness and its implementations using optical fields'',
Optics~Communications (2014), http://dx.doi.org/10.1016/j.optcom.2014.07.012.

\bibitem{QObook}
D.~F.~Walls and G.~J.~Milburn,
\textit{Quantum Optics} (Springer, New York, 1994).

\bibitem{cat1}
J.~S.~Neergaard-Nielsen, B.~M.~Nielsen, C.~Hettich, K.~M\o lmer, and E.~S.~Polzik,
``Generation of a Superposition of Odd Photon Number States for Quantum Information Networks'',
Phys.~Rev.~Lett. \textbf{97}, 083604 (2006).

\bibitem{cat5}
M.~Takeoka, H.~Takahashi, and M.~Sasaki,
``Large-amplitude coherent-state superposition generated by a time-seperated two-photon subtraction from a continuous-wave squeezed vacuum'',
Phys. Rev. A \textbf{77}, 062315 (2008);
M.~Sasaki, M.~Takeoka, and H.~Takahashi,
``Temporally multiplexed superposition states of continuous variables'',
Phys. Rev. A \textbf{77}, 063840 (2008).

\bibitem{cat4}
H.~Takahashi, K.~Wakui, S.~Suzuki, M.~Takeoka, K.~Hayasaka, A.~Furusawa, and M.~Sasaki,
``Generation of Large-Amplitude Coherent-State Superposition via Ancilla-Assisted Photon Subtraction'',
Phys. Rev. Lett. \textbf{101}, 233605 (2008).

\bibitem{cat6}
T.~Gerrits, S.~Glancy, T.~S.~Clement, B.~Calkins, A.~E.~Lita, A.~J.~Miller, A.~L.~Migdall, S.~W.~Nam, R.~P.~Mirin and E.~Knill,
``Generation of optical coherent-state superpositions by number-resolving photon subtraction form the squeezed vacuum'',
Phys. Rev. A \textbf{82}, 031802(R) (2010).

\bibitem{deco}
W.~H.~Zurek,
``Decoherence, einselection, and the quantum origins of the classical'',
Rev. Mod. Phys. \textbf{75}, 715-775 (2003).

\bibitem{Reply}
H.~Jeong, M.~Kang, C.-W.~Lee,
``Reply to Comment on ``Quantification of Macroscopic Quantum Superpositions within Phase Space'''',
http://arxiv.org/abs/1108.0212

\bibitem{Comment}
J.~Gong,
``Comment on ``Quantification of Macroscopic Quantum Superpositions within Phase Space'''',
http://arxiv.org/abs/1106.0062

\bibitem{Sek2012}
P.~Sekatski, N.~Sangouard, M.~Stobi\'{n}ska, F. Bussi\`{e}res, M. Afzelius, and N. Gisin,
``Proposal for exploring macroscopic entanglement with a single photon and coherent states'',
Phys. Rev. A \textbf{86}, 060301(R) (2012).

\bibitem{Lvov2012}
A.~I.~Lvovsky, R.~Ghobadi, A.~Chandra, A.~S.~Prasad, and C.~Simon,
``Observation of micro-macro entanglement of light'',
Nature Physics \textbf{9}, 541-544 (2013).

\bibitem{Bruno2012}
N.~Bruno, A.~Martin, P.~Sekatski, N.~Sangouard, R.~T.~Thew, and N.~Gisin,
``Displacement of entanglement back and forth between the micro and macro domains'',
Nature Physics. \textbf{9}, 545-548 (2013).

\bibitem{Louisell}
W.~H.~Louisell,
\textit{Quantum Statistical Properties of Radiation} (Wile, New York, 1973).

\bibitem{WM2009}
H.~M.~Wiseman and G.~J.~Milburn,
\textit{Quantum Measurement and Control} (Cambridge University Press, 2009).

\bibitem{swap2}
A.~K.~Ekert, C.~M.~Alves, D.~K.~L.~Oi, M.~Horodecki, P.~Horodecki and L.~C.~Kwek,
``Direct Estimations of Linear and Nonlinear Functionals of a Quantum State'',
Phys. Rev. Lett. \textbf{88}, 217901 (2002).

\bibitem{swap1}
R.~Filip,
``Overlap and entanglement-witness measurements'',
Phys. Rev. A \textbf{65}, 062320 (2002).

\bibitem{swap3}
P.~Horodecki and A.~Ekert,
``Method for Direct Detection of Quantum Entanglement''
Phys. Rev. Lett. \textbf{89}, 127902 (2002).

\bibitem{swap4}
M.~Hendrych, M.~Du\v{s}ek, R.~Filip, and J.~Fiur\'{a}\v{s}ek,
``Simple optical measurement of the overlap and fidelity of quantum states'',
Phys. Lett. A \textbf{310}, 95-100 (2003).

\bibitem{swap5}
C.~M.~Alves, P.~Horodecki, D.~K.~L.~Oi, L.~C.~Kwek and A.~K.~Ekert
 ``Direct estimation of functionals of density operators by local operations and classical communication'',
Phys. Rev. A \textbf{68}, 032306 (2003).

\bibitem{swap6}
C.~M.~Alves and D.~Jaksch,
``Multipartite Entanglement Detection in Bosons'',
Phys. Rev. Lett. \textbf{93}, 110501 (2004).

\bibitem{swap7}
K.~L.~Pregnell,
``Measuring Nonlinear Functionals of Quantum Harmonic Oscillator States'',
Phys. Rev. Lett. \textbf{96}, 060501 (2006).

\bibitem{added_01}
H.~Nakazato, T.~Tanaka, K.~Yuasa, G.~Florio, and S.~Pascazio,
``Measurement scheme for purity based on two two-body gates'',
Phys. Rev. A \textbf{85}, 042316 (2012).

\bibitem{added_02}
T.~Tanaka, G.~Kimura, and H.~Nakazato,
``Possibility of a minimal purity-measurement scheme critically depends on the parity of dimension of the quantum system''
Phys. Rev. A \textbf{87}, 012303 (2013).

\bibitem{SML2013}
S.~M.~Lee, S.-K.~Choi, and H.~S.~Park
 ``Experimental direct estimation of nonlinear functionls of photonic quantum states via interferometry with a controlled-swap operation'',
Optics Express \textbf{21}, 17824-17830 (2013).

\bibitem{Leonhardt}
U.~Leonhardt,
``Quantum statistics of a lossless beam splitter: SU(2) symmetry in phase space'',
Phys. Rev. A \textbf{48}, 3265-3277 (1993).

\bibitem{kim1995}
M.~S.~Kim and N.~Imoto,
``Phase-sensitive reservoir modeled by beam splitters'',
Phys. Rev. A \textbf{52}, 2401-2410 (1995).

\bibitem{Daley}
A.~J.~Daley, H.~Pichler, J.~Schachenmayer and P.~Zoller,
``Measuring Entanglement Growth in Quench Dynamics of Bosons in an Optical Lattice'',
Phys. Rev. Lett. \textbf{109}, 020505 (2012).

\bibitem{ZQ2011}
X.-Q.~Zhou, T.~C.~Ralph, P.~Kalasuwan, M.~Zhang, A.~Peruzzo, B.~P.~Lanyon, and J.~L.~O'Brien,
``Adding control to arbitrary unknown quanutm operations'',
Nature Communications \textbf{2}, 413 (2011).

\bibitem{resolving2}
B.~Calkins, P.~L.~Mennea, A.~E.~Lita, B.~J.~Metcalf, W.~S.~Kolthammer, A.~Lamas-Linares, J.~B.~Spring, P.~C.~Humphreys, R.~P.~Mirin, J.~C.~Gates, P.~G.~R.~Smith, I.~A.~Walmsley, T.~Gerrits and S.~W.~Nam,
``High quantum-efficiency photon-number-resolving detector for photonic on-chip information processing'',
Optics Express, \textbf{21}, 22657-22670 (2013).

\bibitem{pc1}
H.~Weinfurter,
``Experimental Bell-state Analysis'',
Europhys. Lett. \textbf{25}, 559-564 (1994).

\bibitem{pc2}
S.~L.~Braunstein and A.~Mann,
``Measurement of the Bell operator and quantum teleportation'',
Phys. Rev. A \textbf{51}, R1727-R1730 (1995)

\bibitem{pc4}
K.~Mattle, H.~Weinfurter, P.~G.~Kwiat, A.~Zeilinger,
``Dense Coding in Experimental Quantum Communication'',
Phys. Rev. Lett. {\bf 76},  4656-4659 (1996).

\bibitem{qnd}
V.~B.~Braginsky,
``Classical and quantum restrictions on the detection of weak disturbances of a macroscopic oscillator'',
Zh. Eksp. Teor. Fiz. \textbf{53}, 1434-1441 (1967).

\bibitem{Brune}
M.~Brune, E.~Hagley, J.~Dreyer, X.~Ma\^itre, A.~Maali, C.~Wunderlich, J.~M.~Raimond, and S.~Haroche,
``Observing the Progressive Decoherence of the ``Meter'' in a Quantum Measurement'',
Phys. Rev. Lett. \textbf{77}, 4887-4890 (1996).

\bibitem{Nogues}
G.~Nogues, A.~Rauschenbeutel, S.~Osnaghi, M.~Brune, J.~M.~Raimond, and S.~Haroche,
``Seeing a single photon without destroying it'',
Nature \textbf{400}, 239-242 (1999).

\bibitem{Filho}
R.~L.~de Matos Filho and W.~Vogel,
``Quantum Nondemolition Measurement of the Motional Energy of a Trapped Atom'',
Phys. Rev. Lett. \textbf{76}, 4520-4523 (1996).

\bibitem{Davidovich}
L.~Davidovich, M.~Orszag, and N.~Zagury,
``Quantum nondemolition measurements of vibrational populations in ionic traps'',
Phys. Rev. A \textbf{54}, 5118-5125 (1996).

\bibitem{Imoto}
N.~Imoto, H.~A.~Haus, and Y.~Yamamoto,
``Quantum nondemolition measurement of the photon number via the optical Kerr effect'',
Phys. Rev. A \textbf{32}, 2287-2292 (1985).

\bibitem{Greentree}
A.~D.~Greentree, R.~G.~Beausoleil, L.~C.~L.~Hollenberg, W.~J.~Munro, K.~Nemoto, S.~Prawer, and T.~P.~Spiller,
``Single photon quantum non-demolition measurements in the presence of inhomogeneous broadening'',
New J. Phys. \textbf{11}, 093005 (2009).

\bibitem{Gerry}
C.~C.~Gerry,
``Generation of Schr\"{o}dinger cats and entangled coherent states in the motion of a trapped ionby a dispersive interaction'',
Phys. Rev. A \textbf{55}, 2478-2481 (1997).

\bibitem{cnot1}
E.~Knill, R.~Laflamme and G.~J.~Milburn,
``A scheme for efficient quantum computation with linear optics'',
Nature \textbf{409}, 46-52 (2001).

\bibitem{cnot2}
J.~L.~O'Brien, G.~J.~Pryde, A.~G.~White, T.~C.~Ralph and D.~Branning,
``Demonstration of an all-optical quantum controlled-NOT gate'',
Nature \textbf{426}, 264-267 (2003).

\bibitem{mix1}
H.~Jeong and T.~C.~Ralph,
`` Transfer of Nonclassical Properties from a Microscopic Superposition to Macroscopic Thermal States in the High Temperature Limit'',
Phys. Rev. Lett. \textbf{97}, 100401 (2006);

\bibitem{mix1-2}
H.~Jeong and T.~C.~Ralph,
``Quantum superpositions and entanglement of thermal states at high temperatures and their applications to quantum-information processing'',
Phys. Rev. A \textbf{76}, 042103 (2007).

\bibitem{Nha2008}
H.~Jeong, J.~Lee, and H.~Nha,
``Decoherece of highly mixed macroscopic quantum superpositions''
J. Opt. Soc. Am. B, \text{25}, 1025-1030 (2008).

\bibitem{mix2}
H.~Jeong, M.~Paternostro, and T.~C.~Ralph,
``Failure of Local Realism Revealed by Extremely-Coarse-Grained Measurements'',
Phys. Rev. Lett. \textbf{102}, 060403 (2009).

\bibitem{spincat}
J.-S.~Lee and A.~K.~Khitrin,
``Twelve-spin ``Schr\"{o}dinger cats'''',
Appl. Phys. Lett. \textbf{87}, 204109 (2005).

\bibitem{tomo1}
D.~F.~V.~James, P.~G.~Kwiat, W.~J.~Munro, and A.~G.~White,
``Measurement of qubits'',
Phys. Rev. A \textbf{64}, 052312 (2001).

\bibitem{tomo2}
J.~B.~Altepeter, E.~R.~Jeffrey and P.~G.~Kwiat, 
``Photonic State Tomography'', 
Advances In Atomic, Molecular, and Optical Physics \textbf{52}, 105-159 (2005).

\bibitem{tomo3}
A.~I.~Lvovsky and M.~G.~Raymer,
``Continuous-variable optical quantum-state tomography'',
Rev. Mod. Phys. \textbf{81}, 299-332 (2009).



\end{thebibliography}
\end{document}